\documentclass[12pt]{article}

\usepackage{amssymb}
\usepackage{amsmath}
\usepackage[dvips]{graphicx}

\newcommand{\cF}{\mathcal{F}}
\newcommand{\la}{\langle}
\newcommand{\ra}{\rangle}
\newcommand{\rar}{\rightarrow}
\newcommand{\cala}{\mathcal{A}}
\newcommand{\calb}{\mathcal{B}}
\newcommand{\dd}{\mathrm{d}}
\newcommand{\we}{\wedge}

\newcommand{\al}{\alpha}
\newcommand{\de}{\delta}

\newcommand{\G}{\Gamma}
\newcommand{\ka}{\kappa}
\newcommand{\e}{\epsilon}
\newcommand{\te}{\theta}

\newcommand{\Fslash}{\hspace{0.1cm}\ensuremath \raisebox{0.025cm}{\slash}\hspace{-0.3cm} \mathcal{F}}

\def\be{\begin{equation}}
\def\ee{\end{equation}}
\def\bea{\begin{eqnarray}}
\def\eea{\end{eqnarray}}

\begin{document}

\title{Baryonic Condensates on the Conifold}
\author{Marcus K. Benna,  Anatoly Dymarsky and Igor R. Klebanov}
\maketitle
\begin{center}
\textit{Department of Physics and
Princeton Center for Theoretical Physics,\\ Princeton
University, Princeton, NJ  08544}
\end{center}

\vspace{-9cm}
\begin{flushright}
hep-th/0612136\\
PUPT-2220\\
ITEP-TH-34/06
\end{flushright}
\vspace{7cm}

\abstract{We provide new evidence for the gauge/string duality
between the baryonic branch of the
cascading $SU(k(M+1))\times SU(kM)$ gauge theory and a family of type IIB
flux backgrounds based on warped products of the deformed conifold
and $\mathbb{R}^{3,1}$.
We show that a Euclidean D5-brane wrapping all six deformed conifold directions
can be used to measure the baryon expectation values, and present arguments based on
$\ka$-symmetry and the equations of motion that identify the gauge bundles required to
ensure worldvolume supersymmetry of
this object. Furthermore, we investigate its coupling to the
pseudoscalar and scalar modes associated with the phase and
magnitude, respectively, of the baryon expectation value. We find
that these massless modes perturb the Dirac-Born-Infeld and
Chern-Simons terms of the D5-brane action in a way consistent with
our identification of the baryonic condensates. We match the
scaling dimension of the baryon operators computed from the D5-brane action
with that found in the cascading gauge theory.
We also derive and numerically evaluate an expression that describes the variation of the baryon
expectation values along the supergravity dual of the baryonic branch.
}

\newpage

\section{Introduction}

Consideration of a stack of $N$ D3-branes leads to
the conjectured
duality of $\mathcal{N}=4$ super Yang-Mills theory to type IIB string theory
on $AdS_5 \times S^5$
\cite{Maldacena:1997re,Gubser:1998bc,Witten:1998qj}.
A different, $\mathcal{N}=1$ supersymmetric example of the AdS/CFT correspondence
follows from placing the stack of D3-branes at the tip of the conifold
\cite{Klebanov:1998hh,MP}. This suggests a duality between a certain
$SU(N)\times SU(N)$ superconformal gauge theory and type IIB string theory on
$AdS_5\times T^{1,1}$. Addition of $M$ D5-branes wrapped over the
two-sphere near the tip of the conifold changes the gauge group to
$SU(N+M)\times SU(N)$ \cite{Gubser:1998fp,KN}. This theory is non-conformal; it
undergoes a cascade of Seiberg dualities \cite{Seiberg:1994pq} $\mathrm{SU}(N+M)\times\mathrm{SU}(N)
\rightarrow \mathrm{SU}(N-M)\times\mathrm{SU}(N)$ as it flows from the
UV to the IR \cite{KT,KS} (for reviews, see
\cite{Herzog:2001xk,Strassler:2005qs}).

The gauge theory contains two doublets of bifundamental,
chiral superfields $A_i, B_j$ (with $i,j=1,2$). In the conformal case,
$M=0$, it has continuous global
symmetries $\mathrm{SU}(2)_ A\times\mathrm{SU}(2)_B\times\mathrm{U}(1)_R\times
\mathrm{U}(1)_B$.
The two $\mathrm{SU}(2)$ groups rotate the doublets $A_i$ and $B_j$,
while one $\mathrm{U}(1)$ is an R-symmetry.
The remaining $\mathrm{U}(1)$ factor corresponds to the baryon number
symmetry which we will be most interested in.
As argued in \cite{KS,Aharony,GHK,DKS}, in the cascading theory where $N$
is an integer multiple of $M$, $N=kM$,
 this symmetry is spontaneously broken
by condensates of baryonic operators. In this paper we will provide a
quantitative verification of this effect.

For $N=kM$ the last step of the cascade is
an $\mathrm{SU}(2M)\times\mathrm{SU}(M)$ theory which admits
two baryon operators (sometimes referred to as baryon and antibaryon)
\begin{eqnarray} \label{baryonops}
\mathcal{A} &\sim& \epsilon_{\al_1\al_2\ldots\al_{2M}}(A_1)_1^{\al_1}(A_1)_2^{\al_2}\ldots(A_1)_M^{\al_M}
(A_2)_1^{\al_{M+1}}(A_2)_2^{\al_{M+2}}\ldots(A_1)_M^{\al_{2M}}\ , \nonumber \\
\mathcal{B} &\sim& \epsilon_{\al_1\al_2\ldots\al_{2M}}(B_1)_1^{\al_1}(B_1)_2^{\al_2}\ldots(B_1)_M^{\al_M}
(B_2)_1^{\al_{M+1}}(B_2)_2^{\al_{M+2}}\ldots(B_1)_M^{\al_{2M}}\ .\qquad
\end{eqnarray}
Baryon operators of the general $\mathrm{SU}(M(k+1))\times\mathrm{SU}(Mk)$
theory have the schematic
form $(A_1A_2)^{k(k+1)M/2}$ and $(B_1B_2)^{k(k+1)M/2}$,
with appropriate contractions described in \cite{Aharony}.
Unlike the ``dibaryon'' operators
of the conformal $\mathrm{SU}(N)\times\mathrm{SU}(N)$
theory \cite{Gubser:1998fp}, $\mathcal{A}$ and $\mathcal{B}$
are singlets under the
two global $\mathrm{SU}(2)$ symmetries.
These operators acquire expectation values that spontaneously break
the $\mathrm{U}(1)_B$ baryon number symmetry; this is why the
gauge theory is said to be on the baryonic branch of its moduli space \cite{Seiberg:1994bz}.
Supersymmetric vacua on the one complex dimensional baryonic branch
are subject to the constraint $\mathcal{A}\mathcal{B} = - \Lambda^{4M}_{2M}$,
and thus we can parameterize it as follows
\begin{equation} \label{ZetaMod}
\mathcal{A} = i\zeta \Lambda^{2M}_{2M}\ ,\qquad
\mathcal{B} = \frac{i}{\zeta} \Lambda^{2M}_{2M}\ .
\end{equation}

The non-singular supergravity dual of the theory with $|\zeta|=1$ is the
warped deformed conifold found in \cite{KS}. In \cite{GHK} the linearized
scalar and pseudoscalar perturbations, corresponding to small deviations
of $\zeta$ from $1$, were constructed. The full set of first-order
equations necessary to describe the entire moduli space of supergravity
backgrounds dual to the baryonic branch, sometimes called the
resolved warped deformed conifolds, was derived and solved numerically in \cite{Butti} (for
a further discussion of the solutions, see \cite{DKS}).

The construction of this moduli space of supergravity backgrounds, which
have just the right symmetries to be identified with the baryonic branch
in the cascading gauge theory, provides an excellent check on the
gauge/string duality in this intricate setting. Yet, one question remains:
how do we identify the baryonic expectation values on the
string side of this duality? Among other things, this is needed to
construct a map between the parameter $U$ that labels the supergravity
solutions, and the parameter $|\zeta|$ in the gauge theory.

The dual string theory
description of the baryon operators (\ref{baryonops}) was first considered
by Aharony \cite{Aharony}. He argued that the heavy ``particle'' dual to such
an operator is described at large $r$ by a D5-brane wrapped over the $T^{1,1}$,
with some D3-branes dissolved in it (to account for this, the
world volume gauge field needs to be turned on). To calculate the two-point function
of baryon operators inserted at $x_1$ and $x_2$ we may use
a semi-classical approach to the AdS/CFT correspondence.
Then we need a (Euclidean) D5-brane
whose world volume has two $T^{1,1}$ boundaries at large $r$, located at $x_1$ and $x_2$.
In this paper we will be interested in a simpler embedding of the D5-brane:
as suggested by Witten \cite{WittenUnp},
the object needed to calculate the baryonic expectation values
is the Euclidean D5-brane that has the appearance of a pointlike instanton
from the four-dimensional point of view, and wraps the remaining six (generalized Calabi-Yau)
directions of the ten-dimensional spacetime. This object has a single
$T^{1,1}$ boundary at
large $r$, corresponding to insertion
of just one baryon operator. As we will find, supersymmetry requires that the
world volume gauge field is also turned on, so there are D3-branes dissolved in
the D5.
This identification will be corroborated by demonstrating that the D5-brane
couples correctly to the pseudoscalar
zero-mode of the theory that changes the phase of the
baryon expectation value \cite{GHK}.

Close to the boundary, a field $\phi$ dual to an operator of dimension
$\Delta$ in the AdS/CFT correspondence behaves as
\begin{equation}
\phi(x,r) = \phi_0(x)\, r^{\Delta-4} + A_\phi(x)\, r^{-\Delta}\ ,
\end{equation}
Here $A_\phi$ is the operator expectation value \cite{KW2}, and
$\phi_0$ is the source for it. In the cascading theory, which is
near-AdS in the UV, the same formulae hold modulo powers of $\ln r$
\cite{Aharony:2005zr,Aharony:2006ce}. The field corresponding to a
baryon will be identified, at a semi-classical level, with
$e^{-S_{D5}(r)}$, where $S_{D5}(r)$ is the action of a D5-brane
wrapping the Calabi-Yau coordinates up to the radial coordinate
cut-off $r$. The different baryon operators ${\mathcal A},
\overline{{\mathcal A}}, {\mathcal B}, \overline{{\mathcal B}}$ will
be distinguished by the two possible D5-brane orientations, and the
two possible $\kappa$-symmetric choices for the world volume gauge
field that has to be turned on inside the D5-brane. In the cascading
gauge theory there is no source added for baryonic operators, hence
we find that $\phi_0=0$. On the other hand, the term scaling as
$r^{-\Delta}$ is indeed revealed by our calculation of
$e^{-S_{D5}(r)}$ as a function of the radial cut-off, allowing us to
find the dimensions of the baryon operators, and the values of their
condensates.

This paper is structured as follows. In the remainder of section 1
we review the geometry of the deformed conifold, and the warped
supergravity backgrounds dual to the baryonic branch, including the
corresponding Killing spinors. We also review the $\kappa$-symmetry
conditions for D-brane embeddings, and briefly discuss a number of
brane configurations that satisfy them. Section 2 is devoted to the
derivation of the first-order equation for the gauge field. We first
discuss a Lorentzian D7-brane wrapping the warped deformed conifold
directions, before presenting a parallel treatment for the more
subtle case of the Euclidean D5-brane wrapping the conifold. Section
3 is devoted to the physics of the D5-instanton in the KS
background. From the behavior of the D5-brane action as a function
of the radial cut-off we extract the dimension of the baryon
operator, and show that it matches the expectations from the dual
cascading gauge theory. We also show that the D5-brane couples to
the baryonic branch complex modulus in the way consistent with our
identification of the condensates. In particular, we demonstrate
that pseudoscalar perturbations of the backgrounds shift the phase
of the baryon expectation value. We generalize to the complete
baryonic branch in section 4 where we compute the baryon expectation
values as a function of the supergravity modulus $U$. The product of
the expectation values calculated from the D5-brane action is shown
to be independent of $U$ in agreement with (\ref{ZetaMod}). Finally,
we present an integral expression for their ratio and evaluate it
numerically, which provides a relation between the baryonic branch
modulus $|\zeta|$ in the gauge theory and the modulus $U$ in the
dual supergravity description, and show that they satisfy
$\mathcal{A}\mathcal{B}= {\rm const}$. We conclude briefly in
section~5.

\subsection{Review of Warped Deformed Conifolds }

We start our discussion with a review of the warped deformed conifold (KS) background
\cite{KS}, which is dual to a locus on the baryonic branch where
$|{\cal A}|= |{\cal B}|$.
Then we review the generalization of the background to the entire baryonic branch found by
Butti et.~al.~\cite{Butti}.

The warped deformed conifold is a warped product of four-dimensional flat space and
an $SU(2)\times SU(2)$ Calabi-Yau three-fold $\mathcal{M}$:
\begin{eqnarray}
ds^2=h(t)^{-1/2}dx_{3,1}^2+h(t)^{1/2}ds_\mathcal{M}^2\ .
\end{eqnarray}
The deformed conifold $\mathcal{M}$ is described in complex coordinates by the equation
\begin{eqnarray}
\sum_{i=1}^4 z_i^2 = \varepsilon^2\ .
\end{eqnarray}
The warp factor is given by
\begin{eqnarray}
h(t) &=& (g_s M \al')^2 2^{2/3} \varepsilon^{-8/3} I(t)\ ,\\
I(t) &\equiv& 2^{1/3}\int_{t}^\infty dx \frac{x \coth (x) - 1}{\sinh^2 (x)}(\sinh (x) \cosh (x) - x)^{1/3}\ ,
\end{eqnarray}
In the asymptotic near-AdS region,
the radial coordinate $t$ is related to the standard coordinate $r$ by
\begin{equation} \label{randt}
r^2 = \frac{3}{2^{5/3}} \varepsilon^{4/3} e^{2\,t/3}\ .
\end{equation}
Since $\mathcal{M}$ has a topology of
$S^2\times S^3\times \mathbb{R}^+$ it is convenient to introduce the following
one-forms $e_i$ on $S^2$
\begin{eqnarray}
e_1\equiv d\theta_1\ ,\qquad e_2\equiv  - \sin\theta_1 d\phi_1\ ,
\end{eqnarray}
and a set of invariant forms on $S^3$
\begin{eqnarray}
\epsilon_1 &\equiv&\sin\psi\sin\theta_2 d\phi_2+\cos\psi d\theta_2\ , \\
\epsilon_2 &\equiv&   \cos\psi\sin\theta_2 d\phi_2 - \sin\psi d\theta_2\ ,\\
\epsilon_3 &\equiv& d\psi + \cos\theta_2 d\phi_2\ .
\end{eqnarray}
In term of these we define one-forms
\begin{eqnarray}
g_1 &\equiv& {e_2-\epsilon_2\over \sqrt{2}}\ , \quad g_2 \equiv {e_1-\epsilon_1\over \sqrt{2}}\ ,\\
g_3 &\equiv& {e_2+\epsilon_2\over \sqrt{2}}\ , \quad g_4 \equiv {e_1+\epsilon_1\over \sqrt{2}}\ ,\\
g_5 &\equiv& \epsilon_3+\cos\theta_1d\phi_1\ ,
\end{eqnarray}
which allow for a concise description of the Calabi-Yau metric on $\mathcal{M}$:
\begin{equation}
ds^2_\mathcal{M}={\varepsilon^{4/3} K(t)\over 2}\left[\sinh^2\left({t\over 2}\right)
\left(g_1^2+g_2^2\right)+\cosh^2\left({t\over 2}\right)
\left(g_3^2+g_4^2\right)+{1\over 3K(t)^3}\left(dt^2+g_5^2\right)\right] \ ,
\end{equation}
where
\begin{equation}
K(t) \equiv { \left(\sinh(t)\cosh(t)-t\right)^{1/3}\over \sinh(t)}\ .
\end{equation}
The dilaton $\phi$ is constant, but there are non-trivial three- and
five-form fluxes in this background \cite{KS}. The NS-NS two-form is given by
\begin{equation}
B_2 = \frac{g_s M\al'}{2}\frac{t \coth(t) - 1}{\sinh(t)}
\left[\sinh^2\left(\frac{t}{2}\right)\, g^1\we g^2 + \cosh^2\left(\frac{t}{2}\right)\, g^3\we g^4 \right],
\end{equation}
and the R-R fluxes are most compactly written as
\begin{eqnarray} \label{fluxcompact}
F_3 &=& \frac{M\al'}{2}\left\{g^3\we g^4 \we g^5 +
d \left[\frac{\sinh(t) - t}{2 \sinh(t)}(g^1\we g^3+g^2\we g^4) \right] \right\}, \\ \label{fluxcompact2}
\tilde{F}_5 &=& dC_4 + B_2 \we F_3 = (1+\ast)\,(B_2 \we F_3)\ .
\end{eqnarray}
Corresponding R-R potentials are easily found:
\begin{eqnarray}
C_2 &=& \frac{M \al'}{2} \big[\frac{\psi}{2}(g^1\we g^2 + g^3\we g^4)-\frac{1}{2}\cos\te_1\cos\te_2\, d\phi_1\we d\phi_2 \nonumber \\  &&\qquad\quad - \frac{t}{2 \sinh(t)}(g^1\we g^3+g^2\we g^4)\big]\ , \\
C_4 &=& \frac{1}{g_s h(t)}\,dx^0\we dx^1 \we dx^2 \we dx^3\ .
\end{eqnarray}
From here on we set the deformation parameter $\varepsilon$ to unity for
notational simplicity, and also choose $M \al' = 2$ and $g_s = 1$.

The KS solution is invariant under the $\mathbb{Z}_2$ symmetry ${\cal I}$, which exchanges $(\theta_1,\phi_1)$ with $(\theta_2,\phi_2)$
accompanied by the action of $-I$ of SL(2, $\mathbb{Z}$).
On the gauge theory side, this symmetry exchanges the $\mathcal{A}$ and
$\mathcal{B}$ baryons.
Therefore, the KS solution corresponds to $|\zeta|=1$ in (\ref{ZetaMod}).
There is a continuous family of solutions which generalize KS and break this
${\cal I}$--symmetry \cite{GHK,Butti}.
This family is dual to the entire baryonic branch of the cascading gauge theory,
parameterized by $\zeta$ (only the modulus of $\zeta$ is manifest in these backgrounds).
The corresponding metric can be written in the form of the Papadopoulos-Tseytlin
 ansatz \cite{PT}
in the string frame:
\begin{eqnarray}
ds^2 =  e^{2A}  dx^2_{3,1} + e^x ds_\mathcal{M}^2  = e^{2A}  dx^2_{3,1} + \sum_{i=1}^6 G_i^2\ ,
\end{eqnarray}
where
\begin{eqnarray} \label{Gforms}
G_1\equiv e^{(x+g)/2}\,e_1\ ,&& \qquad
G_2\equiv {\cosh(t)+a\over \sinh(t)}\,e^{(x+g)/2}\,e_2 + {e^g \over \sinh(t)}\,e^{(x-g)/2}\,(\epsilon_2-a e_2)\ ,\nonumber\\
G_3\equiv e^{(x-g)/2}\,(\epsilon_1-a e_1)\ ,&&\qquad
G_4\equiv {e^g \over \sinh(t)}\,e^{(x+g)/2}\,e_2 - {\cosh(t)+a\over \sinh(t)}\,e^{(x-g)/2}\,(\epsilon_2-a e_2)\ ,\nonumber\\
G_5\equiv e^{x/2}\,v^{-1/2}dt\ ,&& \qquad
G_6\equiv e^{x/2}\,v^{-1/2}g_5\ .
\end{eqnarray}
While in the KS case there was a single warp factor $h(t)$, now we
find several functions $ A(t), x(t), g(t), a(t), v(t)$.

In terms of these one-forms the Calabi-Yau $(3,0)$ form is
\begin{equation}
\Omega = (G_1 + i G_2)\we (G_3 + i G_4)\we (G_5 + i G_6)\ ,
\end{equation}
and the fundamental $(1,1)$ form is
\begin{equation}
J ={i\over 2} \Big [ (G_1 + i G_2) \we (G_1- i G_2) + (G_3 + i G_4) \we (G_3- i G_4)
+ (G_5 + i G_6) \we (G_5- i G_6) \Big ]\ .
\end{equation}
The background also contains the fluxes
\begin{eqnarray}
B_2 &=&
 h_1\,(\epsilon_1 \we \epsilon_2 + e_1 \we e_2)
+ \chi\,(e_1 \we e_2 - \epsilon_1 \we \epsilon_2)  + h_2\, (\epsilon_1 \we e_2 -  \epsilon_2 \we e_1 )
\ , \nonumber\\
F_3 &=& -{1\over 2} g_5\we \big[  \epsilon_1 \we \epsilon_2  +  e_1 \we e_2 -
b\,  (\epsilon_1 \we e_2 - \epsilon_2 \we e_1) \big]
-{1\over 2}\,  dt \we \big[ b'\, (\epsilon_1 \we e_1 +
\epsilon_2 \we e_2) \big]\ ,\nonumber\\
\tilde{F}_5 &=& {\cal F}_5  +  *_{10}{\cal F}_5\ ,\qquad
{\cal F}_5 = -(h_1 + b h_2)\, e_1 \we e_2 \we \epsilon_1 \we \epsilon_2 \we \epsilon_3\ ,
\end{eqnarray}
parameterized by functions
$h_1(t), h_2(t), b(t)$ and $\chi(t)$. In addition, the dilaton $\phi$ now also depends on the
radial coordinate $t$.

The functions $a$ and $v$ satisfy a system of coupled first order differential equations
\cite{Butti} whose solutions are known in closed form only in the KS \cite{KS}
and the Chamseddine-Volkov-Maldacena-Nunez (CVMN) \cite{CV,MN} limits.
All other functions
$A,x,g,h_1,h_2,b,\chi,\phi$ are unambiguously determined by $a$ and $v$ through the relations
\begin{eqnarray}
e^{-4A} = U^{-2}\left(e^{-2\phi}-1\right)\ ,&&\qquad e^{2x}={(bC-1)^2\over 4(aC-1)^2}
e^{2g+2\phi}(1-e^{2\phi})\ ,\\
e^{2g}=-1-a^2 + 2a\,C\ ,&&\qquad h_1 = -h_2\,C\ ,\\
h_2= {e^{2\phi}(bC-1)\over 2S}\ ,&&\qquad  b={t\over S}\ ,\\
\chi'=a(b-C)(aC-1)e^{2(\phi-g)}\ ,&&\qquad \phi'= {\left( C-b\right) {\left(
a\,C-1 \right) }^2 \over \left( b\,C-1 \right) S}\,e^{-2\,g}\ ,
\end{eqnarray}
where $C \equiv -\cosh(t),\ S \equiv -\sinh(t)$,
and we remind the reader that we set ${M\alpha'/2} = \varepsilon = g_s =1$,
and require $\phi(\infty)=0$. In writing these equations we have specialized to the baryonic branch by imposing appropriate
boundary conditions at infinity \cite{DKS}; namely $\eta = 1$ in the notation of \cite{Butti}.
Varying $\eta$ produces
a more general, two parameter family of SU(3) structure backgrounds,
 that also include the CVMN solution \cite{CV,MN}, which requires $\eta=0$ \cite{Butti}.
The baryonic branch ($\eta=1$) family of supergravity solutions is labelled by one real
``resolution parameter'' $U$ \cite{DKS}.
While the leading asymptotics of all supergravity backgrounds dual to the baryonic
branch are given by the KT solution \cite{KT}, terms subleading at large $t$ depend on~$U$.
This family of supergravity
solutions preserves the $SU(2)\times SU(2)$ symmetry, but for $U\neq 0$ breaks
the $\mathbb{Z}_2$ symmetry ${\cal I}$ of the KS background.

On the baryonic branch we can consider a transformation that takes
$\zeta$ into $\zeta^{-1}$, or equivalently
$U$ into $-U$. This transformation leaves $v$ invariant and changes $a$ as follows
\begin{eqnarray}\label{Utrans}
a\rightarrow -{a\over 1+2a\cosh(t)}\ .
\end{eqnarray}
It is straightforward to check that  $ae^{-g}$ is invariant
while $(1+a\cosh(t))e^{-g}$ changes sign.
This transformation also exchanges $e^g+a^2e^{-g}$ with $e^{-g}$ and therefore it is equivalent
to the exchange of $(\theta_1,\phi_1)$ and  $(\theta_2,\phi_2)$
involved in the ${\cal I}$--symmetry.

\subsection{D-Branes, $\kappa$-Symmetry and Killing Spinors of the Conifold \label{killing}}

A Dirichlet $p$-brane
(with $p$ spatially extended dimensions)
in string theory is
described by an action consisting of two terms
\cite{Polchinski,Bachas:1998rg,Johnson:2000ch}:
the Dirac-Born-Infeld action, which is essentially a minimal area action including non-linear
electrodynamics, and the Chern-Simons action, which describes the coupling to the R-R background fields:
\begin{equation}\label{DBICS}
S = S_{DBI}+S_{CS} = {\bf -}\int_\mathcal{W}
\mathrm{d}^{p+1}\sigma e^{-\phi} \sqrt{-\det(G + \mathcal{F})} +
\int_\mathcal{W} e^{\mathcal{F}} \wedge C \ .
\end{equation}
Here $\mathcal{W}$ is the worldvolume of the brane and we have set
the brane tension to unity. Further, $G$ is the induced metric on
the worldvolume, $\mathcal{F} = F_2 + B_2$ is the sum of the gauge field strength
$F_2 = dA_1$ and the pullback  of the NS-NS two-form field, and $C =
\sum_i C_i$ is the formal sum of the R-R potentials. In
superstring theory all these fields should really be understood as
superfields, but we shall ignore fermionic excitations here.

Wick rotation of this action to Euclidian space such that all
$p+1$ directions become spatially extended (which leads to a
Euclidean worldvolume D-instanton) effectively multiplies the action by a factor of $i$.
This cancels the minus sign under the square root in the DBI term
and leaves it real since the determinant is now positive. The
CS term however is purely imaginary now. Consequently the equations of motion
that follow from the DBI and CS terms now have be satisfied
independently of each other if we insist on the gauge field being real.

The action (\ref{DBICS}) is invariant on shell under the so-called
$\kappa$-symmetry \cite{BT,BT2,CederwallEtAl}. This allows us to find first-order
equations for supersymmetric configurations which are easier to solve than the
second order equations of motion. The $\kappa$-symmetry condition can be written as
\begin{eqnarray}
\Gamma_{\ka} \e=\e\ ,
\end{eqnarray}
where $\e$ is a doublet of Majorana-Weyl spinors, and the
operator $\Gamma_{\ka}$ is specified below.
Satisfying this equation guarantees worldvolume supersymmetry in the probe brane
approximation, and every solution for which $\e$ is a Killing spinor
corresponds to a supersymmetry compatible with those preserved by the background.

The decomposition of a Weyl spinor $\e$ into a doublet of Majorana-Weyl spinors
\begin{eqnarray}
\e=\left(\begin{array}{c}
  \e_1 \\
  \e_2 \\
\end{array}\right)
\end{eqnarray}
is achieved by projecting onto the eigenstates of charge conjugation\footnote{Given any spinor
$\e$ we denote its charge conjugate by $\e^*$, which of course is represented by
complex conjugation and left multiplication by a charge conjugation matrix $B$.
We do not write $B$ explicitly here, though its presence is understood.}
$\e_1 = (\e + \e^*)/2$ and $\e_2 = (\e - \e^*)/2i$.

In IIB superstring theory on a $(9,1)$ signature spacetime,
the $\kappa$-symmetry operator $\Gamma_{\ka}$ for
a Lorentzian D-brane extended along the time direction $x_0$ and $p$ spatial
directions is given by
\begin{eqnarray}\label{GammaKappa}
\G_\ka= \frac{\sqrt{-\det G}}{\sqrt{-\det(G+\cF)}}\sum_{n=0}^{\infty}  (-1)^n \Fslash^n \Gamma_{(p+1)} \otimes (\sigma_3)^{n+\frac{p-3}{2}} i \sigma_2\ ,\\
\Gamma_{(p+1)}\equiv \frac{1}{(p+1)!\sqrt{-\det G}}\, \e^{\,\mu_1 \ldots \mu_{p+1}} \Gamma_{\mu_1 \ldots \mu_{p+1}} \ ,\\
\Fslash^n \equiv \frac{1}{2^n n!} \Gamma_{\nu_1 \ldots \nu_{2n}}\cF_{\sigma_1 \sigma_2} \ldots \cF_{\sigma_{2n-1}\sigma_{2n}} G^{\nu_1 \sigma_1} \ldots G^{\nu_{2n}\sigma_{2n}}\ .
\end{eqnarray}
Here $\sigma_i$ are the usual Pauli matrices. We use Greek labels for the worldvolume
indices of the D-brane and consequentially the $\Gamma_\mu$ are induced Dirac matrices.
In what follows we denote the Minkowski spacetime coordinates by $x_0 \ldots x_3$
and label the tangent space of the internal manifold $\mathcal{M}$ by $1, 2 \ldots 6$
in reference to the basis one-forms (\ref{Gforms}).
The expression for $\Gamma_{\ka}$ can be significantly simplified for an embedding covering
all six directions of the deformed conifold, in which case we simply
align the worldvolume tangent space with that of $\mathcal{M}$.

The Killing spinor $\Psi$ of the supergravity backgrounds dual to
the baryonic branch is built out of a six-dimensional pure spinor
$\eta^-$ and an arbitrary spinor $\zeta^-$ of negative four-dimensional
chirality,
\bea\label{KillingSpinor}
\Psi = \al\, \zeta^-\otimes \eta^- + i\beta\, \zeta^+\otimes \eta^+\ ,\\
\label{pures}
(\Gamma_1-i\Gamma_2)\zeta^- \otimes\eta^-=(\Gamma_3-i\Gamma_4)\zeta^-\otimes\eta^-=
(\Gamma_5-i\Gamma_6)\zeta^-\otimes\eta^-=0\ ,
\eea
where $\eta^+ = (\eta^-)^\ast$ and
$\zeta^+ = (\zeta^-)^\ast$.
The functions $\alpha$ and $\beta$
are real \cite{Butti,DKS} and given by
\begin{eqnarray}
\alpha={e^{\phi/4}(1+e^{\phi})^{3/8}\over(1-e^{\phi})^{1/8}}\
,\qquad
 \beta= {e^{\phi/4}(1-e^{\phi})^{3/8}\over(1+e^{\phi})^{1/8}}\ .
\end{eqnarray}
(this expression for $\beta$
is for $U>0$; $\beta$ changes sign when $U$ does).
The corresponding Majorana-Weyl spinors $\Psi_1$ and $\Psi_2$ are
\begin{eqnarray}
\Psi_1={1\over 2}\left((\alpha-i\beta)\zeta^-\otimes\eta^- +(\alpha +i \beta)\zeta^+\otimes\eta^+\right)\ ,\\
\Psi_2={1\over 2i}\left((\alpha+i\beta)\zeta^-\otimes\eta^-
-(\alpha -i \beta)\zeta^+\otimes\eta^+\right)\ .
\end{eqnarray}

\subsection{Branes Wrapping the Angular Directions}

In the context of the conifold, the closest analogue to the baryon vertex in
AdS$_5\times \mathrm{S}^5$ that was discussed in
\cite{WittenBaryons,Imamura:1998gk,Callan:1998iq},
would be a D5-brane wrapping the five angular directions of the internal space,
with worldvolume coordinates $\sigma^\mu = (x^0, \te_1, \phi_1, \te_2, \phi_2, \psi)$.
The brane describing the baryon vertex in
AdS$_5\times \mathrm{S}^5$ has ``BI-on'' spikes corresponding to
fundamental strings attached to the brane and ending on the boundary of AdS,
indicating that it is not a gauge-invariant object.
Here however, we are interested in gauge-invariant, supersymmetric objects,
that are candidate duals to
chiral operators in the gauge theory,
so we might try to consider a smooth embedding at constant radial coordinate
(the difference between a ``baryon" and a ``baryon vertex" was already stressed in \cite{WittenBaryons}).

To avoid having the BI-on spikes, it was proposed \cite{Aharony} that we should use an
appropriate combination of D5-branes wrapping all the angular coordinates, and of D3-branes
wrapping the $S^3$. This is equivalent to turning on a particular gauge field on the wrapped
D5-brane. Unfortunately, it is not clear how to maintain the supersymmetry
of such an object.
It is not hard to see, for example from the appropriate $\ka$-symmetry equations,
that a (Lorentzian) D5-brane wrapping the five angular direction of the conifold
and embedded at constant $r$ cannot be a supersymmetric object.
The $\ka$-symmetry equation seems to call for an additional constraint of the form
$\G_{x^0\psi} \e^\ast = -i \e$ on the Killing spinors,
which would imply also $\G_{x^0 r} \e^\ast = - \e$, i.e. precisely what we would expect
for strings stretched in the radial direction.
However, such a projection does not commute with the other conditions that the Killing spinors have to satisfy and thus is not consistent.
This was pointed out in \cite{Arean} for the case of the singular conifold
\cite{Klebanov:1998hh}, and the argument carries over to the deformed conifold.
Even with a worldvolume gauge field
such a D5-brane cannot be a BPS object.

The same conclusion also follows from the equation of motion for the radial
component of the embedding $X^M(\zeta^{\mu})$. The leading term (as $r \rar \infty$)
in the D-brane Lagrangian arises from the $B_2$-field contribution to the DBI term and is
proportional to $r\,(\ln r)^2$, so this brane is bound to contract and move to smaller $r$,
until eventually it reaches the tip of the conifold, where the two-cycle collapses
and the brane unwraps.

On the other hand, as suggested by Aharony \cite{Aharony},
the D5-branes with D3-branes dissolved within them
are the ``particles'' dual to
the baryon operators.
As suggested by Witten \cite{WittenUnp}, to find the baryonic condensates
we need to consider a {\it Euclidean} D5-brane wrapping
the deformed conifold directions, with a certain gauge field turned on.
While there are no non-trivial two-cycles in this case, the worldvolume gauge field
does modify the coupling of this D-instanton to the R-R potential $C_4$.
We will show that such a configuration can be made
$\ka$-symmetric and then yields the baryonic condensates consistent with the
gauge theory expectations.

As a first example of a supersymmetric brane wrapping all the
angular directions, we shall discuss a D7-brane wrapping the warped
deformed conifold, with the remaining one space and one time
directions extended in $\mathbb{R}^{3,1}$. The supersymmetry
conditions for general D-branes in ${\cal N}=1$ backgrounds were
derived in \cite{Marino:1999af,Martucci:2005ht,Martucci:2006ij}, and
our results will be consistent with theirs. We will show that the
Lorentzian D7-brane configuration on the KS background is
supersymmetric in the absence of a worldvolume gauge-field, though
the $\ka$-symmetry analysis will also reveal supersymmetric
configurations with non-zero gauge field. The fact that switching on
this field is not required for supersymmetry might have been guessed
from a naive counting argument. This embedding of the D7-brane
should be mutually supersymmetric with the D3-branes filling the
$\mathbb{R}^{3,1}$, since the number of Neumann-Dirichlet directions
for strings stretched between them equals eight.

The object we are most interested in is
the Euclidean D5-brane completely wrapped on the conifold. In contrast to the case of the D7-brane,
we will find that supersymmetry requires a non-trivial gauge field on the worldvolume.
Again this is consistent with the naive count of Neumann-Dirichlet directions with the D3-branes,
which gives ten in this case and thus indicates that these branes cannot be mutually
supersymmetric if $F_2 =0$.

\section{Derivation of the First-Order Equation for
the Worldvolume Gauge Bundle} \label{Derivations}

In this section we derive the first-order equation of motion that
the U(1) gauge field has to satisfy to obtain a supersymmetric
configuration. Because the $\kappa$-symmetry of the Euclidean
D5-brane is subtle, we will first discuss the closely related case
of a Lorentzian D7-brane wrapping the six-dimensional deformed
conifold, with non-zero gauge bundle only in these directions. This
object is extended as a string in the $\mathbb{R}^{3,1}$ but in the
case of a non-compact space dual to the cascading gauge theory the
tension of such a string diverges with the cut-off as $e^{2t/3}$.
Therefore, this string is not part of the gauge theory spectrum.

\subsection{$\kappa$-Symmetry of the Lorentzian D7-Brane}

The explicit form of the $\ka$-symmetry equation for the D7 brane with non-trivial U(1) bundle
on the six-dimensional internal space is given by
\begin{equation}\label{D7Kappa}
\binom{\e_1}{\e_2} = \G_\ka \binom{\e_1}{\e_2} \sim
\left[-( \Fslash + \Fslash^3)\,\sigma_3 + (1 + \Fslash^2) \right] i\sigma_2\,\G_{x^0x^1123456}
\binom{\e_1}{\e_2}\ ,
\end{equation}
For the case of Euclidean D-branes wrapping certain cycles in
Calabi-Yau manifolds, it was shown in \cite{Marino:1999af} that the
$\ka$-symmetry condition (\ref{D7Kappa}) can be rewritten in more
geometrical terms. This results in the conditions that
$\mathcal{F}^{2,0} = 0$, and that
\begin{equation}\label{GeomKappaCond}
{1\over2!}J\we J \we \mathcal{F} - {1\over3!} \mathcal{F} \we \mathcal{F} \we \mathcal{F} = \mathfrak{g} \left({1\over3!}J \we J \we J - {1\over2!}J\we \mathcal{F} \we \mathcal{F} \right).
\end{equation}
The constant $\mathfrak{g}$ was found \cite{Marino:1999af}
to encode some information about the geometry,
namely a relative phase between coefficients of the covariantly constant spinors
in the expansion of the $\e_{i}$ \cite{Marino:1999af}.
As we shall see below, the same equation holds in our case of a generalized Calabi-Yau
with fluxes, except that $\mathfrak{g}$ becomes coordinate dependent.

With the
$SU(2)\times SU(2)$ invariant ansatz for the gauge potential
\begin{equation} \label{gaugeans}
A_1 = \xi(t) g_5\ ,
\end{equation}
we find that the gauge-invariant two-form field strength is given by
\begin{eqnarray}
\cF &=& {i e^{-x}\over 2\sinh(t)} \times \\
&&\bigg[e^{-g}\left[\tilde{\xi}(\cosh(t)+2a+a^2\cosh(t)) +
h_2 \sinh^2(t)(1-a^2)\right] (G_1+iG_2)\wedge(G_1-iG_2) \nonumber \\
&&+\,\, e^{g}\left[\tilde{\xi}\cosh(t) - h_2 \sinh^2(t)\right]
(G_3+iG_4)\wedge(G_3-iG_4) \nonumber \\
&&+\,\, \xi' v \sinh(t) (G_5+iG_6)\wedge (G_5-iG_6) +\left[\tilde{\xi}(1+a\cosh(t))- h_2 a \sinh^2(t)\right] \nonumber \\
&&\big((G_1+iG_2)\wedge (G_3-iG_4)+(G_3+iG_4)\wedge (G_1-iG_2)\big)\bigg]\ , \nonumber
\end{eqnarray}
where $\tilde{\xi}=\xi+\chi$. This explicitly shows that $\cF$ is a
$(1,1)$ form, which is one of the $\ka$-symmetry conditions
\cite{Marino:1999af,Martucci:2005ht,Martucci:2006ij}. Now it is
convenient to define
\begin{eqnarray} \label{aandb}
\mathfrak{a}(\xi,t) &\equiv& e^{-2x}[e^{2x} + h_2^2 \sinh^2(t)-(\xi+\chi)^2]\ ,\nonumber \\
\mathfrak{b}(\xi,t) &\equiv& 2 e^{-x-g}\sinh(t) [a(\xi+\chi) - h_2(1+a\cosh(t))]\ .
\end{eqnarray}
In terms of these expressions we find that
\begin{eqnarray}\label{JandF}
{1\over3!} J \we J \we J - {1\over2!} J \we \mathcal{F} \we \mathcal{F} &=& (\mathfrak{a} + v e^{-x} \mathfrak{b}\,\xi')\,\mathrm{vol}_6\ ,\nonumber\\
{1\over2!} J \we J \we \mathcal{F} - {1\over3!} \mathcal{F} \we \mathcal{F} \we \mathcal{F} &=& (- \mathfrak{b} + v e^{-x} \mathfrak{a}\,\xi')\,\mathrm{vol}_6\ ,
\end{eqnarray}
where $\mathrm{vol}_6 = (J \we J \we J) / 3!$. Thus (\ref{GeomKappaCond}) would
lead to a differential equation of the form
\begin{equation} \label{xiprime}
\xi' = {e^x(\mathfrak{g} \mathfrak{a} + \mathfrak{b})
\over v (\mathfrak{a} - \mathfrak{g} \mathfrak{b})}\ ,
\end{equation}
for some as yet undetermined $\mathfrak{g}$. In order to confirm
the validity of this equation and determine the function
$\mathfrak{g}$ we return to the full $\ka$-symmetry equation
(\ref{D7Kappa}) with the Majorana-Weyl spinors $\e_1 = (\Psi +
\Psi^*)/2$ and $\e_2 = (\Psi - \Psi^*)/2i$ constructed from the
Killing spinor. The analysis of this equation is much simplified
by noting that $\Gamma_{1..6}\eta^{\pm}= \mp i\eta^{\pm}$ and that
the spinors $\eta^{\pm}$ are in fact eigenspinors\footnote{For simplicity we drop the four-dimensional spinors $\zeta^\pm$ in $\zeta^\pm\otimes \eta^\pm$.} of $\Fslash ^n$

\begin{eqnarray}\label{}
\Fslash \eta^\pm &=& \pm i\eta^\pm\left(\cF_{12}+\cF_{34}+\cF_{56}\right)\ ,\\
\Fslash^2 \eta^\pm &=& - \eta^\pm \left(\cF_{12}\cF_{34}+\cF_{14}\cF_{23}+\cF_{12}\cF_{56}+\cF_{34}\cF_{56}\right)\ ,\\
\Fslash^3 \eta^\pm &=& \mp i\eta^\pm
\left(\cF_{12}\cF_{34}\cF_{56}+\cF_{14}\cF_{23}\cF_{56}\right)\ ,
\end{eqnarray}
where the indices refer the basis one-forms (\ref{Gforms}). Then it follows from
(\ref{JandF}) that the two terms in the $\ka$-symmetry equation
act on the spinors in a rather simple fashion:
 \begin{eqnarray}
\left[1+\Fslash^2\right]\eta^\pm &=& \left[\mathfrak{a}+ve^{-x}\mathfrak{b}\xi'\right]\eta^\pm\  ,\nonumber\\
\left[\Fslash+\Fslash^3\right]\eta^\pm &=& \pm
i\left[-\mathfrak{b}+ve^{-x}\mathfrak{a}\xi'\right]\eta^\pm\ .
\end{eqnarray}
Using these relations it is easy to see that the Killing spinor
(\ref{KillingSpinor}) indeed solves (\ref{D7Kappa}) provided we
impose the conditions that its four-dimensional parts $\zeta^\pm$
 obey the condition $\G_{x^0x^1}\zeta^\pm\otimes \eta^\pm = \zeta^\pm\otimes \eta^\pm$, and that the gauge field $\xi(t)$ satisfies
 (\ref{xiprime}) with
\begin{equation}
\mathfrak{g}(t) = \mathfrak{g}_7(t) \equiv - {2\al\beta \over
\al^2 - \beta^2}= -e^{-\phi}\sqrt{1-e^{2\phi}}\ .
\end{equation}
Thus indeed (\ref{GeomKappaCond}) holds and (\ref{xiprime}) is the
correct first order differential equation given this function
$\mathfrak{g}(t)$.

The fact that the $\ka$-symmetry condition (\ref{D7Kappa}) is
satisfied implies worldvolume supersymmetry in the probe brane
approximation. However, we also ask for the worldvolume
supersymmetries to be compatible with those of the background. In
order to check how many supersymmetries of the background are
preserved by the brane we need to enumerate the solutions of
(\ref{D7Kappa}) for which $\e_1 + i\e_2$ is not just any spinor,
but a Killing spinor. For the particular case of the D7-brane with
U(1) gauge bundle determined by the first-order equation
(\ref{xiprime}) we saw that Killing spinors of the form
(\ref{KillingSpinor}) solve the $\ka$-symmetry equation if
$\G_{x^0x^1}\zeta^\pm\otimes \eta^\pm = \zeta^\pm\otimes \eta^\pm$, and thus half of the
supersymmetries of the background are preserved.

\subsection{An Equivalent Derivation Starting from the Equation of Motion}

Here we present an alternative derivation of the first-order equations for
the gauge field $\xi(t)$, starting from
the second-order equation of motion. This method has the advantage that it applies
equally well to
Lorentzian D7 and Euclidean D5-branes wrapping the conifold.
The $\kappa$-symmetry argument we employed in the previous section for the D7-brane is
somewhat complicated in the case of the D5-instanton by the fact that we are
forced to Wick rotate to Euclidean spacetime signature where there are no Majorana-Weyl spinors.
However, knowing that a first-order differential equation for the gauge field exists,
as well as
its general features, it is not hard to derive it directly from the
second-order equation of motion.

Since with Euclidean signature the DBI action is real and the CS action pure imaginary,
two sets of equations of motion have to be satisfied simultaneously if we insist
on the gauge field being real.
With the ansatz (\ref{gaugeans})
for the gauge potential, the CS equations are automatically satisfied,
as are five of the DBI equations; only the one for the $g_5$ component of the gauge field
(or equivalently its $\psi$ component) is non-trivial.

In terms of the (implicitly $U$-dependent) functions defined in \cite{Butti}
the determinant that appears in the DBI action is given by
\begin{eqnarray} \label{detGFB}
\mathrm{det}_{\mathcal{M}}(G+\cF) &=& v^{-2} e^{6x}(1+(\xi')^2 v^2 e^{-2x}) \Bigg[1 + e^{-4x}\left((\xi+\chi)^2-\sinh^2(t) h_2^2\right)^2 \nonumber \\
&&- 2e^{-2x}\left((\xi+\chi)^2+\sinh^2(t) h_2^2\right)
\left(1-2 e^{-2g} a^2\sinh^2(t)\right) \nonumber \\
&&- 8e^{-2x -2g}\sinh^2(t)a h_2(\xi+\chi)(1+a\cosh(t)) \Bigg]\ ,
\end{eqnarray}
where we have omitted the angular dependence $\sim \sin^2 \theta_1 \sin^2 \theta_2$.
Here we have only taken into account the six-dimensional internal manifold $\mathcal{M}$.
If the brane is also extended in the Minkowski directions
(but carries zero gauge bundle in these directions) there are additional $\xi$-independent
factors multiplying the DBI determinant that appears in the action (\ref{DBICS}).
E.g.~for the Lorentzian D7-brane this factor is equal to $e^{4A}$.
Using the definitions (\ref{aandb}), the term in square brackets in (\ref{detGFB})
can be written as a sum of squares
$\mathfrak{a}^2 + \mathfrak{b}^2$.

We know from the form of the $\kappa$-symmetry equation that the first-order
differential equation we are looking for must

i) be polynomial (of at most third order) in $\xi$ and its first derivative,

ii) contain $\xi'$ only at linear order (i.e. no $(\xi')^2$ terms),

iii) be such that the determinant factorizes.

In particular the last condition means that when we eliminate $\xi'$ from the action,
the $\xi$-dependent term must be a perfect square, else the factor of $\sqrt{\mathrm{det}_{\mathcal{M}}(G+\cF)}$ in the denominator of (\ref{GammaKappa}) cannot be cancelled by the numerator to give unit eigenvalue. This implies that we must have
\begin{equation}
(1+(\xi')^2 v^2 e^{-2x}) = {\mathfrak{a}^2 + \mathfrak{b}^2 \over \mathfrak{f}^2(\xi,t)}\ ,
\end{equation}
for some $\mathfrak{f}(\xi,t)$, so that
\begin{equation}
\xi' = {e^x\sqrt{\mathfrak{a}^2 + \mathfrak{b}^2 - \mathfrak{f}^2(\xi,t)} \over v\, \mathfrak{f}(\xi,t)}\ .
\end{equation}
Because we expect the equation to be polynomial in $\xi$ one must be able to explicitly take the square root, and thus $\mathfrak{f}(\xi,t)$ can be written as
\begin{equation}
\mathfrak{f}(\xi,t) = {\mathfrak{a} - \mathfrak{g}(t)\mathfrak{b} \over \sqrt{1+\mathfrak{g}^2(t)}}\ ,
\end{equation}
for some function $\mathfrak{g}(t)$, where all the $\xi$ dependence is now implicit in $\mathfrak{a}$ and $\mathfrak{b}$\ . With this ansatz we have
\begin{equation} \label{xiprime2}
\xi' = {e^x(\mathfrak{g} \mathfrak{a} + \mathfrak{b})  \over v (\mathfrak{a} - \mathfrak{g} \mathfrak{b})}\ ,
\end{equation}
which is of the same form as the first order differential equation we derived for the D7-brane in the previous section. The function $\mathfrak{g}$ follows by varying the action with respect to $\xi$ and substituting for $\xi'$ using (\ref{xiprime2}).
It is not difficult to check that the equations of motion that follow from the DBI action of the D7-brane $\int e^{2A-\phi}\sqrt{\mathrm{det}_{\mathcal{M}}(G+\cF)}$ are indeed implied by the first order equation (\ref{xiprime2}) with
\begin{equation} \label{g7}
\mathfrak{g} = \mathfrak{g}_7 = {e^{x-g}(1+a\cosh(t)) \over h_2\sinh(t)} = -e^{-\phi}\sqrt{1-e^{2\phi}}\ ,
\end{equation}
as we found above using a $\ka$-symmetry argument.

Using the same method, we can now find the first-order equation for
the gauge field on the Euclidean D5-brane. Having constrained the equation we are looking for to the form (\ref{xiprime2}) we vary the DBI action $\int e^{-\phi}\sqrt{\mathrm{det}_{\mathcal{M}}(G+\cF)}$ using (\ref{detGFB}) and eliminate $\xi'$ to obtain
\begin{eqnarray}
&& {\delta \over \delta \xi} \left[ e^{-\phi}\sqrt{\mathrm{det}(G+F+B)} \right] =  0 = \nonumber \\
&& {2e^{-\phi}e^{2x}\sqrt{1+\mathfrak{g}^2} \over v (\mathfrak{a} - \mathfrak{g} \mathfrak{b})} \left[-(\xi+\chi)e^{-x}\mathfrak{a} + e^{-g} a \sinh(t) \mathfrak{b} \right]
 - {d \over dt}\left[{e^{-\phi}e^{2x}(\mathfrak{g} \mathfrak{a} + \mathfrak{b}) \over \sqrt{1 + \mathfrak{g}^2}}\right]\ .
\end{eqnarray}
Collecting powers of $\xi$ and equating their coefficients to zero we find differential equations for $\mathfrak{g}(t)$ which are solved simultaneously by
\begin{equation} \label{g-function}
\mathfrak{g} = \mathfrak{g}_5 \equiv - {e^{-x+g} h_2\sinh(t) \over (1+a\cosh(t))} = {e^{\phi}\over\sqrt{1-e^{2\phi}}}\ .
\end{equation}
Substituting this into (\ref{xiprime2}) the first-order equation we were looking for, written out in full, is
\begin{eqnarray} \label{FirstOrderEq}
\xi' =&&  \Big[-h_2 \sinh(t)e^{2g}[e^{2x} + h_2^2 \sinh^2(t)-(\xi+\chi)^2] \nonumber\\
&& + 2e^{2x}\sinh(t)(1+a\cosh(t))[a(\xi+\chi) - h_2(1+a\cosh(t))]\Big] \times \nonumber \\
&& \Big[v e^g \big[(1+a\cosh(t))[e^{2x} + h_2^2 \sinh^2(t)-(\xi+\chi)^2] \nonumber \\
&&+ 2 h_2 \sinh^2(t)[a(\xi+\chi) - h_2(1+a\cosh(t))]\big] \Big]^{-1} \ .
\end{eqnarray}
In spite of its complicated appearance,
this equation can be integrated and can in fact be solved fairly explicitly.
In the KS limit it reduces to a simpler equation (\ref{D5KSFirstOrder})
that will be discussed in section 3.

Let us note here the interesting fact that the Euclidean D5-brane and the Lorentzian
D7-brane are related by $\mathfrak{g}_5 = - 1/\mathfrak{g}_7$. For the D7-brane we
find $\mathfrak{g}_7=0$ for
the KS background (since there $1+a\cosh(t)=0$), while $\mathfrak{g}_7$ diverges
far along the baryonic branch where $h_2 \to 0$, and correspondingly for $\mathfrak{g}_5$
the situation is the other way around\footnote{As a curious aside
note that taking $\mathfrak{g} = 0$ in (\ref{xiprime2}) leads to an equation consistent
with the action $\int e^{2A-2\phi}\sqrt{\mathrm{det}_{\mathcal{M}}(G+\cF)}$. This coincides
with the D7 brane case for the KS solution (since here $\phi=0$), but in general it
is not clear what (if anything) this corresponds to.}.

The first order equation for the gauge bundle we have derived is in fact more general
than we have made explicit, and when written in the form (\ref{FirstOrderEq})
applies to the whole two-parameter $(\eta, U)$
family of SU(3) structure backgrounds discussed in \cite{Butti}.
The baryonic branch in particular corresponds to the
choice of boundary condition $\eta=1$ at $t=\infty$ in the notation of \cite{Butti},
but the above family of solutions also includes the CVMN background \cite{CV,MN}, which
has the linear dilation boundary condition $\eta = 0$ at infinity.
We discuss some details of the Euclidean D5-instanton on the CVMN background in appendix
\ref{appendixMN}.

\subsection{$\kappa$-Symmetry of the Euclidean D5-Brane}

Let us now reconsider
the Euclidean D5-brane using the $\ka$-symmetry approach.
The $\ka$-symmetry projection operator in \cite{BT,CederwallEtAl} was
derived using the superspace formalism for Lorentzian worldvolume branes in (9,1) signature spacetimes, and thus it is not immediately clear if it is applicable to the case of a Euclidean worldvolume instanton which necessarily has to reside in a (10,0) signature spacetime. For now we shall nevertheless proceed by performing just a naive Wick-rotation of the $\ka$-symmetry projector, which simply introduces a factor $-i$ in (\ref{GammaKappa}) such that $\G_{\ka}^2 = 1$ still holds.

The analog of the $\ka$-symmetry condition (\ref{D7Kappa}) for the Euclidean D5-brane is then given by
\begin{equation}\label{D5Kappa}
\binom{\e_1}{\e_2} = \G_\ka \binom{\e_1}{\e_2} \sim \left[-( \Fslash + \Fslash^3) + (1 + \Fslash^2)\,\sigma_3 \right]\sigma_2\,\G_{123456} \binom{\e_1}{\e_2}\ .
\end{equation}
Re-expressing this in geometrical terms leads to an equation of the same form as (\ref{GeomKappaCond}), but now we expect $\mathfrak{g}(t)$ to be equal to $\mathfrak{g}_5(t)$. Using the same ansatz $A_1 = \xi(t)g_5$ as above it is clear that equations~(\ref{JandF}) and thus (\ref{xiprime}) still hold, and of course $\cF$ is still a (1,1) form. Let us mention in passing that Euclidean D5-branes with gauge bundles satisfying $\cF^{2,0}=0$ also play an important role in topological string theory (see e.g.~\cite{Iqbal:2003ds}).

However, with the gauge bundle we derived in the previous subsection (i.e. with~$\mathfrak{g} = \mathfrak{g}_5 = (\al^2 - \beta^2)/(2\al\beta)$) the $\ka$-symmetry equation (\ref{D5Kappa}) does not have solutions for $\e_1 + i\e_2$ being equal to the Killing spinor (\ref{KillingSpinor}). We can find solutions for other spinors by expanding the $\e_i$ in terms of pure spinors:
\begin{equation}
\e_{i} = x_{i}(t)\,\zeta^-\otimes \eta^- + y_{i}(t)\,\zeta^+\otimes \eta^+\ ,
\end{equation}
where $i=1,2$. We find that with this ansatz (\ref{D5Kappa}) is solved if the coefficients satisfy
\begin{equation}\label{SpinorCoeff}
{x_1 \over x_2} = i{(\al - i \beta)^2\over \al^2 + \beta^2}\ , \qquad {y_1 \over y_2} = i{(\al + i \beta)^2\over \al^2 + \beta^2}\ .
\end{equation}
Thus we have obtained a family of spinors (\ref{SpinorCoeff}) that solves the $\ka$-symmetry equation with the correct gauge bundle,
but this family does not seem to contain the Killing spinor (which differs by a sign in $y_1/y_2$).
This would imply that even though for the gauge field configuration we have found there
is worldvolume supersymmetry in the probe brane approximation, these supersymmetries would not be compatible with those of the background.

We believe that this difficulty
is just an artefact of applying the $\ka$-symmetry operator in a Euclidean spacetime to a
Euclidean worldvolume brane without properly taking into account the subtleties of
Wick-rotating the spinors and the projector itself, and that the D5-instanton
does preserve the background supersymmetries.
In fact it is known that for a Euclidean D5-brane wrapping six internal dimensions the correct $\kappa$-symmetry equations are not the ones obtained by the naive Wick rotation we performed above, but instead are identical to those for a Lorentzian D9-brane\footnote{We would like to thank L. Martucci for pointing this out to us.}. The $\kappa$-symmetry conditions for the Lorentzian D9-brane
lead to equations identical to (\ref{SpinorCoeff}) except for a change of sign on the right hand side of the equation 
for $y_1/y_2$,
so that they are now satisfied by the Killing spinor. This shows that the 
worldvolume gauge field found above is consistent with properly defined 
$\kappa$-symmetry.

In either case we consider the independent derivation of the first-order equation
(\ref{FirstOrderEq}) in the previous subsection a compelling argument that this
gauge bundle is in fact the correct one for our purposes, which will be corroborated below
by the successful extraction of the baryon operator dimension from its large $t$ behaviour.

\section{Euclidean D5-Brane on the KS Background}

We will now specialize
the discussion of the previous section to the case of a Euclidean D5-brane wrapping the
deformed conifold in the KS background. Since this background is known analytically,
the formulae are more explicit in this case.
We interpret the Euclidean D5-brane (which has the appearance of a pointlike instanton in Minkowski space) as the dual of the baryon in the field theory, in the sense that
its action captures information about the (scale-dependent) anomalous dimension of the baryon operator,
as well as its expectation value.

\subsection{The Gauge Field and
the Integrated Form of the Action}

For the KS background, with $a = -1/\cosh(t)$ and $\chi=0$, the first-order differential equation
(\ref{FirstOrderEq}) simplifies to
\begin{equation}
\xi' =  {e^{2x} + h_2^2 \sinh^2(t)-\xi^2 \over 2 v \xi}\ ,
\end{equation}
or more explicitly, substituting in the KS expressions for $x,h_2$ and $v$:
\begin{equation} \label{D5KSFirstOrder}
3 \,{\sinh(t)\cosh(t)-t \over \sinh^2(t)}\, \xi'\xi + \xi^2 = {(\sinh(t)\cosh(t)-t)^{2/3}h \over 16} + {1 \over 4}\,(t\coth(t)-1)^2\ .
\end{equation}
Note that there is no $\xi'\xi^2$ term. For this reason we can multiply the equation by
an integrating factor to turn the left hand side into the total derivative
$[(\sinh(t)\cosh(t)-t)^{1/3}\xi^2]'$ and reduce the equation to the integral
\begin{equation} \label{KSSol}
\xi^2 = (\sinh(t)\cosh(t)-t)^{-1/3}  J(t) \ ,
\end{equation}
where
\begin{equation} \label{Jform}
J(t)=
\int_0^t \left({\sinh^2(x)\,h(x) \over 24} + {\sinh^2(x)(x\coth(x)-1)^2 \over 6\,
(\sinh(x)\cosh(x)-x)^{2/3}}\right) dx\ .
\end{equation}
We have set the integration constant to zero by requiring regularity at $t=0$.
The integral looks ``almost" like the explicitly computable one
\begin{eqnarray}
&&\int_0^t \left({\sinh^2(x)\,h(x) \over 24} + {\sinh^2(x)(x\coth(x)-1)^2 \over 18\,(\sinh(x)\cosh(x)-x)^{2/3}}
\right)dx \nonumber \\
&&= {1\over 48}(\sinh(t)\cosh(t)-t)\, h(t)  + {1 \over 12}\,(t\coth(t)-1)^2 (\sinh(t)\cosh(t)-t)^{1/3}\ ,
\end{eqnarray}
but a relative factor of 3 in the second term of (\ref{Jform}) prevents us from performing
it in closed form.

Now consider the DBI action of the Euclidean D5-brane with this worldvolume gauge field.
Neglecting the five angular integrals for the time being, and focussing on the radial integral,
we see that the Lagrangian is in fact a total derivative, and thus the action is given by
\begin{eqnarray} \label{KSAction}
S_{DBI} &\sim& \int \dd t\, e^{-\phi}\sqrt{\det{G+\cF}} \nonumber \\
&=&
-{1 \over 3(\sinh(t)\cosh(t)-t)^{1/2}} J^{3/2}
\\ \nonumber
&&+ \left[{(\sinh(t)\cosh(t)-t)^{1/2} h \over 16} + {(t\coth(t)-1)^2\over4(\sinh(t)\cosh(t)-t)^{1/6}} \right] J^{1/2}\ .
\end{eqnarray}

We are particularly interested in the UV behaviour of these quantities. From (\ref{KSSol})
it is easy to find the asymptotic expansion of the gauge field as $t \rightarrow \infty$:
\begin{equation}
\xi^2 \rightarrow {1 \over 4}t^2 - {7\over8}t + {47\over32} +
\mathcal{O}(e^{-2t/3})\ .
\end{equation}
Note that to leading order this approximates $h_2^2 \sinh^2(t)$,
so for large $t$ the coefficients of the $F_2$ and $B_2$ fields become equal and
cancellations occur in the action. This is essential for obtaining
the $t^3$ behaviour of the action for large cut-off $t$,
which as we will see gives the correct $t^2$
scaling of the baryon operator dimensions.

To extract the asymptotic behaviour of the action we will use the
integrated form (\ref{KSAction}). The leading terms in the
expansion are easily found analytically, with the result
\bea
\label{SAsymp}
S_{DBI} = \int \dd t e^{-\phi}\sqrt{\mathrm{det}(G+\cF)}
&\rightarrow& {1\over 6}(t^2+t-2) \left({1\over 4}t^2-{7\over 8}t+{47\over32}\right)^{1/2}+\mathcal{O}(e^{-2t/3})\nonumber \\
&\rightarrow& {1 \over 12}t^3 - {1\over 16}t^2 - {25\over128}t + {943\over 1536}+
\mathcal{O}(1/t)\ .
\eea
Below we will argue that the $\mathcal{O}(1)$
term in this expansion determines the
expectation value of the baryon operator. Of particular interest
is the variation of this expectation value along the baryonic
branch; we will investigate it in the next section. First,
however, we will give a field theoretic interpretation to the
terms that increase with $t$. As we will see, the coefficients of
these divergent terms are universal for all backgrounds along the
baryonic branch.

\subsection{Scaling Dimension of Baryon Operator}

We have seen that for large cut-off
$r$ (i.e.~large $t$), the DBI action of the Euclidean D5-brane will behave as $S(r) \sim (\ln(r))^3$.
Since this object corresponds to the baryon in the field theory, we
expect that $\exp(-S)$ is related to  $r^{-\Delta}$, where $\Delta$ is the scaling dimension of the baryon operator.

To make this statement more precise we consider the RG flow equation relating the operator
dimension $\Delta$ to the boundary behavior of the dual field $\phi(r)$:
\begin{equation} \label{RGeq}
-r {\dd \phi(r) \over \dd r} = \Delta(r) \phi(r)\ .
\end{equation}
This equation obviously holds in the usual AdS/CFT case where all operator dimensions
have a limit as the UV cut-off is removed. The case of cascading theories is more subtle, since
there exist operators, such as the baryons, whose dimensions grow in the UV. As we will
see, in these cases (\ref{RGeq}) is still applicable.
Identifying the field dual to a baryon operator as
\begin{equation} \label{dualfield}
\phi (r)\sim \exp(-S(r))
\ ,\end{equation}
 we find
\begin{equation} \label{OpDimAction}
\Delta(r) = r {\dd S(r) \over \dd r} = {\dd S(r) \over \dd \ln(r)}\ .
\end{equation}

To calculate the scaling dimension of the baryon in the gauge theory,
we simply count the number of constituent fields required to build a baryon operator
for a given gauge group $SU(kM)\times SU((k+1)M)$ and multiply by the dimension of the
chiral superfield $A$ or $B$; the latter approaches $3/4$ in the UV where the theory
is quasi-conformal. This gives
\begin{equation} \label{OpDim}
\Delta(r) = {3\over4} M k(k+1) =  {27 g_s^2 M^3 \over 16 \pi^2} (\ln(r))^2 + \mathcal{O}(\ln(r))\ ,
\end{equation}
where $k$ labels the cascade steps and we have used the asymptotic
expression for the radius (energy scale) at which the $k$th Seiberg duality is performed:
\begin{equation}
r_k = r_0 \exp\left({2\pi k \over 3 g_s M}\right).
\end{equation}
Here and in the remainder of this subsection we keep factors of $g_s, M, \varepsilon$ and $\alpha'$ explicit.

Let us now compare this to the scaling dimension we obtain from the action of the D5-instanton
according to eq.~(\ref{OpDimAction}).
The leading term in the action is $t^3/12$, which is multiplied by a factor
$(g_s M \alpha' /2)^3$ that we had previously set to one,
a factor $64 \pi^3$ from the previously neglected five angular integrals and a factor of $\tau_5 = (2\pi)^{-5}\alpha'^{-3}g_s^{-1}$. Therefore, using
(\ref{randt}) we have
\begin{equation}
S = {t^3\over 12} \left({g_s M \alpha' \over 2}\right)^3 {64 \pi^3 \over (2 \pi)^5 \alpha'^3 g_s} +
\mathcal{O}(t^2) =  {9 g_s^2 M^3 \over 16 \pi^2} (\ln(r))^3 + \mathcal{O}((\ln(r))^2)\ .
\end{equation}
{}From (\ref{OpDimAction}) we find that this string theoretic calculation gives
\begin{equation}
\Delta(r)  =  {27 g_s^2 M^3 \over 16 \pi^2} (\ln(r))^2 + \mathcal{O}(\ln(r))\ .
\end{equation}
The term of leading order in $\ln(r)$ is in perfect agreement with
the gauge theory result (\ref{OpDim}).
We consider this a strong argument that the relation (\ref{dualfield})
between the Euclidean D5-brane action
and the field dual to the baryon is indeed correct.
It would be nice
to also compare the terms of order $\ln(r)$
in the operator dimension, but we postpone this more detailed study
to future work.

\subsection{Chern-Simons Action - Coupling to Pseudoscalar Mode and the Phase of the Baryonic Condensate}

Let us now turn to a discussion of the Chern-Simons terms in the D-brane action.
Given our conventions (\ref{fluxcompact2})
for the gauge-invariant and self-dual five-form field strength
$\tilde{F}_5$, there is a slight subtlety in the CS term of the action (\ref{DBICS}).  Its standard
form, given above, is valid with the choice of conventions where
$\tilde{F}_5 = F_5 + H_3 \wedge C_2 = d C_4 + d B_2 \wedge
C_2$. In these conventions $d C_4$ is invariant under $B_2$ gauge
transformations $B_2 \rightarrow B_2 + d\lambda_1$, but transforms
under $C_2$ gauge transformations $C_2 \rightarrow C_2 + d
\Lambda_1$ such as to leave $\tilde{F_5}$ invariant.
However, we work in different conventions where $\tilde{F}_5 = d C_4
+ B_2 \wedge F_3$; here $d C_4$ changes under $B_2$ gauge
transformations. This choice also alters the form of the CS term
in the action. The new R-R fields are obtained by $C_4 \rar C_4
{\bf +} B_2 \wedge C_2$ combined with $C_2 \rar -C_2$ everywhere else, which
modifies some of the terms in the CS action that will be relevant
for us:
\begin{equation}
\frac{1}{2}\int C_2 \wedge \cF \wedge \cF + \int C_4 \wedge \cF \rar -\frac{1}{2}\int C_2 \wedge F \wedge F + \frac{1}{2}\int C_2 \wedge B \wedge B + \int C_4 \wedge \cF\ .
\end{equation}

For the KS background the CS action simply vanishes. However, it is interesting to consider small perturbations around it.
The pseudoscalar glueball discovered in \cite{GHK} is the Goldstone boson of
the broken $U(1)$ baryon number symmetry;
it is associated with the phase of
the baryon expectation value. This massless mode is a
deformation of the R-R fields
(which is generated for example by a D1-string extended in
$\mathbb{R}^{3,1}$) given by
\begin{eqnarray} \label{RRDef}
\de F_3 &=& \ast_4 da + f_2(t)\,
da\wedge dg^5 + f_2'(t)\, da \wedge dt \wedge g^5\ , \nonumber \\
\de \tilde{F}_5 &=& (1+\ast) \de F_3\wedge B_2 = \left(\ast_4 da -
\frac{h(t)}{6 K^2(t)}\, da \wedge dt\wedge g^5\right) \wedge
B_2\ , \qquad
\end{eqnarray}
where $a(x^0,x^1,x^2,x^3)$ is a pseudoscalar field in four dimensions that satisfies
$d\ast_4da=0$ and would experience monodromy around a D-string.
This deformation solves the supergravity equations with
\begin{equation}
f_2(t) = \frac{1}{6\,K^2(t)\sinh^2 t} \int_0^t dx h(x) \sinh^2 (x)\ .
\end{equation}
If we wish to identify the exponential $\exp(-S) = \exp(-S_{DBI}-S_{CS})$
of the brane action
(or more precisely the constant term in its asymptotic expansion as $t \to \infty$)
with the baryon expectation value, then the pseudoscalar massless mode
has to shift the phase of this quantity, contained in the imaginary Chern-Simons term.
The DBI action is obviously unaffected by this deformation of the background since the
NS-NS fields are unchanged. This is consistent with the magnitudes of the baryon expectation values
being unaffected by the pseudoscalar mode; these magnitudes depend only on the scalar
modulus $U$ in supergravity, corresponding to $|\zeta|$ in the gauge theory.

The phase $\exp(-S_{CS})$ by itself is not gauge invariant and
thus not physical. Because our brane configuration has a boundary
at $t = \infty$, only the difference in phase  $\exp(-\Delta S_{CS}) = \exp(-i \Delta\phi)$ between two
Euclidean D5-branes displaced slightly in one of the transverse directions
(i.e. between two instantons at different points in Minkowski
space) is gauge-invariant. Taking into account the anomalous
Bianchi identities for $F_5$ and $F_7$ and the R-R gauge
transformations we see that this gauge-invariant phase difference
is given by
\bea \Delta \phi =\Delta \phi_B +\Delta \phi_F\ ,\eea
where
\bea \Delta\phi_B &=& \int \left[ {1\over2} \delta F_3 \we B
\we B +
\delta F_5 \we B + \ast_{10} \delta F_3 \right]\ ,\\
\label{phiF} \Delta\phi_F &=& \int \left[ -{1\over2} \delta F_3 \we
F \we F + \delta F_5 \we F \right]\ . \eea The integrals are taken
over the six internal dimensions as well as a line in Minkowski
space. Note that here $F_5 = \dd C_4 = \tilde{F_5} - B_2 \we F_3$.
For small perturbations around KS
 the contribution $\Delta\phi_F$ from the coupling to
the gauge field vanishes (the first term in
(\ref{phiF}) is a total derivative with vanishing boundary terms,
while the second term doesn't have the right angular structure to
give a non-zero result). Substituting the explicit form of the R-R
deformations from (\ref{RRDef}) we find that
the phase difference is
\begin{equation}
\Delta\phi_B = -{1\over2}\int \left({h\over6 K^2} + f_2' \right)(t
\coth(t) -1)^2 \,\dd a \we \dd t \we g^1 \we g^2 \we g^3 \we g^4
\we g^5\ .
\end{equation}
We can interpret $\Delta \phi$ as $\Delta a$ times a baryon number. It is satisfying to see that the pseudoscalar Goldstone mode indeed shifts the phase of the baryon expectation value and not its magnitude.
A more stringent test of our interpretation, which we leave for future work,
would be to carry out this computation for the whole baryonic branch and check
whether the numerical value of the baryon number computed this way
is independent of the modulus $U$.
This is rather difficult, since the pseudoscalar mode at a general point
along the baryonic branch is not explicitly known at present.

\section{Euclidean D5-Brane on the Baryonic Branch}

In this section
we extend the discussion of the previous section from the KS solution to the entire
baryonic branch. In particular we are interested in the dependence of the baryon expectation value
on the modulus $U$ of the supergravity solutions.
All supergravity backgrounds dual to
the baryonic branch have the same asymptotics \cite{DKS}
and we will see that the leading terms (cubic, quadratic and linear in $t$) in the asymptotic expansion of the action (\ref{SAsymp}) are universal.
This implies that the leading scaling dimensions of the baryon operators
do not depend on $U$, consistent with field theory expectations.
However, the finite term in the asymptotic expansion of the brane action
does depend on $U$. This provides a map from the
one-parameter family of supergravity solutions labelled
by $U$ to the
family of field theory vacua with different baryon expectation values (\ref{ZetaMod}),
parameterized by $\zeta$.

\subsection{Solving for the Gauge Field and Integrating the Action}

Having derived the differential equation that determines the gauge field in full generality in Section \ref{Derivations}, let us now turn to a more detailed investigation of the first order
equation (\ref{FirstOrderEq}). First of all we note that it can be
rewritten as
\begin{eqnarray}
\label{D5_firstorder} &&{d \over dt}\Bigg[-{1 \over 3} \xi^3 +
\left({a h_2 \sinh^2(t)\over 1+a\cosh(t)} - \chi\right)\xi^2 +
\left(e^{2x} - h_2^2 \sinh^2(t) - \chi^2 + {2 a h_2
\sinh^2(t)\over 1+a\cosh(t)}\chi \right)\xi \Bigg] \nonumber \\
&=& -{h_2 \sinh(t) e^g \over v (1+a\cosh(t))}[e^{2x} + h_2^2
\sinh^2(t) - \chi^2] + {2 e^{2x} \sinh(t) \over v e^g}[a \chi -
h_2(1+a\cosh(t))]\ .
\end{eqnarray}
For notational convenience we define
\begin{eqnarray}
\tilde{\xi} &\equiv& \xi + \chi\ , \\
\mathfrak{A}(t) &\equiv& {a h_2 \sinh^2(t)\over 1+a\cosh(t)}\ , \\
\mathfrak{B}(t) &\equiv& e^{2x} - h_2^2 \sinh^2(t)\ , \\
\label{rho}
\rho(t) &\equiv& \int_0^t\bigg[{h_2 \sinh(t) e^g \over v (1+a\cosh(t))}[e^{2x} + h_2^2 \sinh^2(t)] \nonumber \\
&& + {2 e^{2x} h_2 \sinh(t) (1+a\cosh(t)) \over v e^g} - [e^{2x} -
h_2^2 \sinh^2(t)]\chi'\bigg] dt\ ,
\end{eqnarray}
which allows us to write (\ref{D5_firstorder}) more compactly
\begin{equation} \label{Cubic}
{\dd \over \dd t}\Big[-{1 \over 3}\tilde{\xi}^3 +
\mathfrak{A}(t)\tilde{\xi}^2 + \mathfrak{B}(t)\tilde{\xi} +
\rho(t)\Big] = 0\ .
\end{equation}
Thus the solutions for the shifted
field
$\tilde{\xi}$  are given by the roots of the third order polynomial
\begin{equation} \label{Cubicnew}
-{1 \over 3}\tilde{\xi}^3 +
\mathfrak{A}(t)\tilde{\xi}^2 + \mathfrak{B}(t)\tilde{\xi} +
\rho(t) = C \ ,
\end{equation}
where $C$ is the integration constant.\footnote{
This equation is quite general; it does not assume
boundary conditions $\eta=1$  that characterize the baryonic
branch \cite{DKS}. In particular this result is also valid for a
brane embedded in the CVMN solution \cite{CV,MN}. This case is somewhat off the
main line of this paper, but in appendix \ref{appendixMN} we
briefly summarize results for the CVMN background analogous to those
presented here.}
To fix it, we consider the small $t$ expansion, which is valid for
any $U$
\bea
\mathfrak{A}&\sim& t+\mathcal{O}(t^3)\ ,\\
\mathfrak{B}&\sim& t^2+\mathcal{O}(t^4)\ ,\\
\rho&\sim& t^3+\mathcal{O}(t^4)\ .
 \eea
Note that at $t=0$ all coefficients in (\ref{Cubicnew}) vanish, except
the first one; therefore, the integration constant $C$ has to
be zero for this cubic to admit more than one real solution.
Then we find that $\tilde{\xi}=0$ at $t=0$ for any
solution on the baryonic branch.

Let us examine the cubic equation (\ref{Cubicnew}) more closely in
the KS limit ($U \to 0$) to see how our earlier result
(\ref{KSSol}) is recovered. In the $U\rightarrow 0$ limit
$a\rightarrow -{1\over \cosh(t)}$ and therefore $(1+a\cosh(t))$ vanishes. For small $U$ \cite{GHK,DKS,Butti}
 \bea
(1+a\cosh(t))=2^{-5/3}UZ(t)+\mathcal{O}(U^2)\ , \\
Z(t) \equiv {(t- \tanh(t)) \over (\sinh(t)\cosh(t)-1)^{1/3}}\ .
\eea
In this case $\mathfrak{A}$ and the first term in $\rho$
diverge as $U^{-1}$. All other terms can be dropped
and we have instead of (\ref{Cubic}) \bea \label{ghk}
\tilde{\xi}^2 {a h_2\sinh^2(t)\over Z(t)}+\int_0^t dt\
{h_2\sinh(t)e^{g} \over v Z(t)}[e^{2x}+h_2^2 \sinh(t)^2]=0 \ .
\eea After substituting the KS values for $a,v,h_2,x$ we recover
(\ref{KSSol}).

While it would be desirable to obtain a closed form expression for
the integral $\rho(t)$ in order to evaluate $\xi$ explicitly, this
appears to be impossible, since even in the KS case we cannot  perform
the corresponding integral $J(t)$.

Evaluating the DBI Lagrangian on-shell using (\ref{xiprime2}) we find
\begin{equation} \label{DBILag}
e^{-\phi}\sqrt{\mathrm{det}(G+\cF)} =
{e^{-\phi}e^{3x}\sqrt{1 + \mathfrak{g}^2}\,(\mathfrak{a}^2+\mathfrak{b}^2)
\over v|\mathfrak{a} - \mathfrak{g} \mathfrak{b}|}\ ,
\end{equation}
where we have taken the absolute value  since the sign of
$\mathfrak{a} - \mathfrak{g} \mathfrak{b}$ will turn out to depend
on which root of equation (\ref{Cubicnew}) we pick.

For the baryonic branch backgrounds
we can show that the action is a total derivative.
First note that
the DBI Lagrangian (\ref{DBILag}) can be rewritten
in the form
\begin{eqnarray}
e^{-\phi}\sqrt{\mathrm{det}(G+\cF)} &=&
{e^{-\phi}e^{3x}\over v\sqrt{1+\mathfrak{g}^2}}
{(\mathfrak{g}\mathfrak{a} + \mathfrak{b})^2 + (\mathfrak{a} -
\mathfrak{g}\mathfrak{b})^2 \over |\mathfrak{a} -
\mathfrak{g}\mathfrak{b}|} \nonumber \\
&=& \left|{e^{4x}(1+a\cosh(t))\over v h_2 \sinh(t) e^g} [v e^{-x}
\xi' (\mathfrak{g}\mathfrak{a} + \mathfrak{b}) + (\mathfrak{a} -
\mathfrak{g}\mathfrak{b})]\right|\ , \qquad
\end{eqnarray}
where the right hand side is now cubic in $\xi$ (and its
derivative) much like the differential equation (\ref{xiprime2}).
In fact, substituting for $\mathfrak{a}$, $\mathfrak{b}$
and $\mathfrak{g} = \mathfrak{g}_5$ this equation can be
integrated in the same manner, which results in the action
\bea \label{action_cons}\label{SInt} S=\Big|-{1 \over
3}\tilde{\xi}^3 + \mathfrak{C}(t)\tilde{\xi}^2 +
\mathfrak{D}(t)\tilde{\xi} + \sigma(t)\Big|\ ,\eea
with
$\mathfrak{C},\mathfrak{D},\sigma$ defined as
\bea
\mathfrak{C}&=&-{e^{2x} a\, (1+a\cosh(t))\over h_2 e^{2g}}\ ,\\
\mathfrak{D}&=&[e^{2x} + h_2^2 \sinh^2(t) + 2e^{2x}(1+a\cosh(t))^2e^{-2g}]\ ,\\
\sigma&=&-\int_0^t \bigg[{e^{2x} (1+a\cosh(t))\over v h_2 \sinh(t)
e^{g}}[e^{2x} - h_2^2 \sinh^2(t)]+   \\  &&\qquad\quad [e^{2x} +
h_2^2 \sinh^2(t)+2e^{2x}(1+a\cosh(t))^2e^{-2g}]\chi'\bigg]dt\ .
\eea
Again the $\xi$-independent term is an integral, that we
denoted by $\sigma(t)$. Thus we have a fairly explicit expression
for the action involving two integrals: $\rho(t)$, which appears in
the equation for $\tilde{\xi}$, and $\sigma(t)$.

To conclude this subsection we will demonstrate that the third
solution of (\ref{Cubic}), which is absent (formally divergent for all $t$) in the KS
case (\ref{KSSol}), produces a badly divergent action and is
therefore unacceptable for any point on the branch. Restoring the
$-\tilde{\xi}^3/3$ term in (\ref{ghk}) we see that in the GHK
region $U\rightarrow 0$ the third solution is simply
\bea {\xi}=-{
2^{2/3}3\over U}\left(\cosh(t)\sinh(t)-t\right)^{1/3}+\mathcal{O}(U)\ .
\eea
The value of the Lagrangian in this case is
\begin{equation}
\sqrt{\mathrm{det}(G+\cF)} = {36 \over U^3}\sinh^2(t) +
\mathcal{O}(U^{-2})\ .
\end{equation}
This expression can be used to extract the leading UV asymptotics of the
Lagrangian for any $U$ as the UV behavior is universal for all $U$:
\bea
\sqrt{\mathrm{det}(G+\cF)} \rightarrow  {9 \over U^3}e^{2t}\ .
\eea
Since the action for the third solution diverges
exponentially at large $t$
 it does not seem possible to interpret this
solution as the dual of an operator in the same sense as we do for
the other two solutions.

\subsection{Baryonic Condensates}

We shall now study the D5-brane action (\ref{SInt}) in more detail.
First we develop an asymptotic expansion of the action (\ref{SInt})
as a function of the cut-off. This expansion is useful because the
divergent terms give the scaling dimension of the baryon operator,
while the finite term encodes its expectation value.\footnote{A
systematic procedure for isolating the finite terms is holographic
renormalization \cite{Skenderis:2002wp,Karch:2005ms}. In this paper
we limit ourselves to a more heuristic approach, which we hope can
be justified through a holographic renormalization procedure. We
leave this for future work.} Then we present a perturbative
treatment of small $U$ region followed by a numerical analysis of
the whole baryonic branch. The main result of this section will be
an expression for the expectation value as a function of $U$ which
can be evaluated numerically. This leads to an explicit relation
between the field theory modulus $|\zeta|$ and the string theory
modulus $U$.

To calculate the baryonic condensates we need asymptotic the behavior
of $\mathfrak{A},\mathfrak{B},\rho$ and
$\mathfrak{C},\mathfrak{D}$ for large $t$. Notice that since for any $U$ the
solution approaches the KS solution at large $t$, the
terms divergent at $U=0$ are UV divergent as well:
\bea \label{asymptotic}
\mathfrak{A} \rightarrow {e^{2t/3}\over U}+\mathcal{O}(e^{-2t/3}) \ ,\\
\mathfrak{B} \rightarrow \mathcal{O}(t^2) \ , \\
\rho\rightarrow -{e^{2t/3}\over U}\left({1\over 4}t^2-{7\over 8}t+{47\over32}\right)+ \mathcal{O}(1)\ ,\\
\mathfrak{C}\rightarrow \mathcal{O}(e^{-2t/3})\ ,\\
\mathfrak{D}\rightarrow \left({1\over 4}t^2-{t\over 8}+{5\over
32}\right)+ \mathcal{O}(e^{-4t/3})\ .
\eea From the expansion for
$\mathfrak{A},\mathfrak{B},\rho$ we find that at large $t$ the gauge
field $\tilde{\xi}$ grows linearly with $t$ and approaches the KS
value with exponential precision
\bea \label{asimpt_xi}
\tilde{\xi}(t,U) \rightarrow \pm\left({1\over 4}t^2-{7\over
8}t+{47\over32}\right)^{1/2}+\mathcal{O}(e^{-2t/3})\ . \eea It is crucial
that the dependence on $U$ in (\ref{asimpt_xi}) is exponentially
suppressed.

Since $\mathfrak{C}$ is exponentially small and the leading term in $\mathfrak{D}$ is $U$-independent we can explicitly express the action (\ref{action_cons}) in terms  of $\sigma$:
\bea
\label{spm}
S_{\pm}(U,t)= S_{\rm div} (t) \pm\sigma(U,t)+\mathcal{O}(e^{-2t/3})\ ,
\eea
where the $U$-independent divergent part of the action is given by
\bea
S_{\rm div} (t) &=& {1\over 6}(t^2+t-2) \left({1\over 4}t^2-{7\over 8}t+{47\over32}\right)^{1/2} \ ,
\eea
Note that
\bea
\left|-{1 \over 3}\tilde{\xi}^3 +
\mathfrak{D}(t)\tilde{\xi}\right|&=& S_{\rm div}(t) +\mathcal{O}(e^{-2t/3})\ .
\eea
The
two signs stand for the two well-behaved solutions $\xi(t)$ corresponding
to the two baryons $\mathcal{A}$ and $\mathcal{B}$. As we argued in section 1, the
$\mathcal{I}$-symmetry which exchanges the $\mathcal{A}$ and
$\mathcal{B}$ baryons is equivalent to changing the sign of $U$. Our explicit
expression (\ref{spm}) confirms that
\bea
S_+(U,t)&=&S_-(-U,t)\ ,\\
S_-(U,t)&=&S_+(-U,t)\ ,
\eea
since $\sigma(U,t)$ is antisymmetric in $U$ according to the arguments presented around (\ref{Utrans}).
In order to find the expectation value of the baryons we evaluate
the action (\ref{SInt}) on these solutions and  remove
the  divergence by subtracting the KS value.
The expectation values hence are
given by $\exp[-\lim_{t \to \infty} S_0(\xi_{1,2})]$, where by
$S_0$ we denote the finite part of the action. It is simplest to
work with the product (normalized to the KS value) and ratio of
the expectation values. The former is given by
\begin{equation}
\label{ratio}
{\la \cala \ra \la \calb \ra \over \la \cala \ra_{KS} \la \calb
\ra_{KS}} = \lim_{t \rar \infty} \exp \left[ S_+(U,t)+S_-(U,t)-2S(0,t)\right]\ ,
\end{equation}
where we have used the fact that the two solutions coincide in the KS case, where  $\sigma=0$.
It follows from (\ref{ratio}) that
\bea
\la \cala \ra \la \calb \ra = \la \cala \ra_{KS} \la \calb
\ra_{KS} \ ,
\eea
which corresponds to the constraint $\cala \calb=-\Lambda^{4M}_{2M}$ in
the gauge theory.
The ratio of the baryon condensates is given by
 \begin{equation}
{\la \cala \ra \over \la \calb \ra} = \lim_{t \rar \infty} \exp
\left[ S_+(U,t)-S_-(U,t)\right]=\lim_{t \rar \infty} e^{2  \sigma}\ ,
\end{equation}
or
\bea
\log \la \cala \ra \simeq \lim_{t \rar \infty} \sigma(t)\ .
\eea

Unfortunately we were not able to calculate $\sigma$  analytically, since the
$U$-dependent terms of order $\mathcal{O}(t^n)
\exp(-2t/3)$ in the integrand are significant.
However, we
can evaluate the integral to first order in $U$ for small $U$:
\begin{eqnarray} \label{ExpSlope}
\sigma &=& 2^{-5/3}U \int_0^\infty \Bigg[ {h \sinh^2(t) \over 12 (\sinh(t) \cosh(t) - t)^{2/3}}  \left({h (\sinh(t) \cosh(t) - t)^{2/3} \over 16} - {(t \coth(t) - 1)^2 \over 4}\right)  \nonumber \\
&-& {(t \coth(t) - 1) (\sinh(t) \cosh(t) - t)^{2/3} \over
\sinh^2(t)}  \left({h (\sinh(t) \cosh(t) - t)^{2/3} \over 16} +
{(t \coth(t) - 1)^2 \over 4}\right) \Bigg] \dd t  \nonumber\\
&\simeq&  3.3773U+\mathcal{O}(U^3)\ ,
\end{eqnarray}
and thus obtain the slope of the expectation values in the vicinity
of KS. Even though we lack analytical arguments that would fix the
behavior of the expectation values for large $U$, we can compute the
integral $\sigma(t)$ numerically. Our results for the expectation
value as a function of the modulus are shown in Figure 1. Since $\la
\mathcal{A} \ra \sim \zeta$ this plot provides a mapping from the
SUGRA modulus $U$ to the field theory modulus $\zeta$ (as we
remarked before, careful holographic renormalization is needed to
check this relation).

\begin{figure}
\begin{center}
\includegraphics[width = 0.8\textwidth]{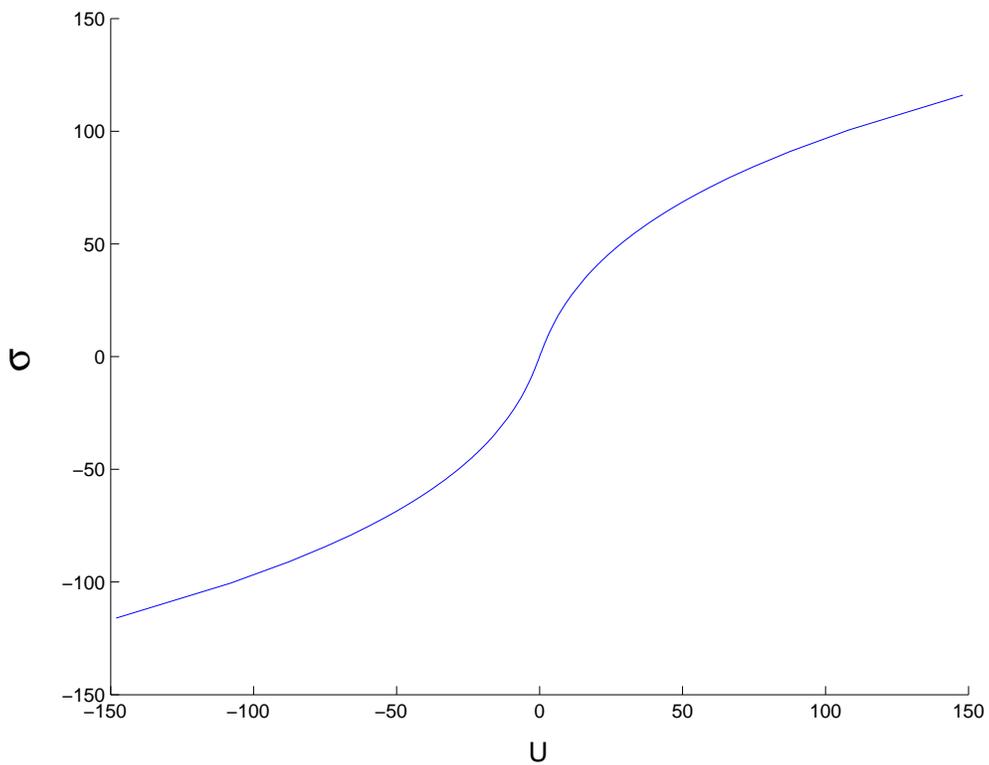}
\end{center}
\caption{Plot of numerical results for the $\mathcal{O}(t^0)$
term in the asymptotic expansion of the action versus $U$.
The slope at $U=0$ matches the value calculated from (\ref{ExpSlope}).
The baryon expectation value $\la \mathcal{A} \ra \sim \la \mathcal{B}
\ra^{-1}$ in units of $\Lambda^{2M}_{2M}$ is given by the exponential of this function.}
\end{figure}

\section{Conclusions}

In previous work, increasingly convincing evidence has been emerging
\cite{KS,Aharony,GHK,DKS}
that the warped deformed conifold background of \cite{KS}
is dual to the cascading gauge theory with condensates of the
baryon operators $\mathcal{A}$ and $\mathcal{B}$.
Furthermore, a one-parameter family of more general warped
deformed conifold backgrounds was constructed \cite{Butti,DKS}
and argued to be dual to the entire baryonic branch of the moduli space,
$\mathcal{A}\mathcal{B}=\mathrm{const.}$

In this paper we present additional, and more direct, evidence for this
identification by calculating the baryonic condensates on the string
theory side of the duality. Following \cite{Aharony,WittenUnp},
we identify the Euclidean D5-branes wrapped over the
deformed conifold, with appropriate gauge fields
turned on, with the fields dual to the baryonic operators in the sense
of gauge/string dualities.
We derive the first order equations for the gauge fields and solve them
explicitly. The solutions  are
subjected to a number of tests. From the behavior of the D5-brane action at large radial cut-off $r$
we deduce the $r$-dependence of the baryon operator dimensions and
match it with that in the cascading gauge theory.
Furthermore, we use the D5-brane action to calculate the condensates
as functions of the modulus $U$ that is explicit in the supergravity
backgrounds. We find that the product of the $\mathcal{A}$ and $\mathcal{B}$
condensates indeed does not depend on $U$.

This calculation also establishes a map between the
parameterizations of the baryonic branch on the string theory and on
the gauge theory sides of the duality. This map should be useful for
comparing other physical quantities along the baryonic branch, and
we hope to return to such comparisons in the future.

\section{Acknowledgements}

We thank E. Witten for the initial suggestion that gave the impetus
to this project, and S. Gubser for useful discussions.
This research was supported
in part by the
National Science Foundation under Grant No. PHY-0243680. The
research of AD is also supported in part by Grant RFBR
04-02-16538, and Grant for Support of Scientific Schools
NSh-8004.2006.2.  Any opinions, findings, and conclusions or
recommendations expressed in this material are those of the
authors and do not necessarily reflect the views of the National
Science Foundation.

\appendix
\section{Appendix: D5-Brane on the CVMN Background} \label{appendixMN}

Here we collect some results for the Euclidean D5-brane in the CVMN background \cite{CV,MN}.
As emphasized in \cite{DKS} and above, this background is not part of the baryonic branch
since its asymptotic behavior at large $t$ is different from
the ``cascading behavior'' found in \cite{KT,KS}. With $h_2=\chi=0$
and after substituting the explicit CVMN expressions \cite{Butti,CV,MN} for the remaining functions,
 the differential equation (\ref{FirstOrderEq}) simplifies to
\begin{equation}
\xi' =  {- t \sinh(t) \xi \over 2 \sqrt{-1+2 t \coth(t)-{t^2 \over \sinh^2(t)}}} \left[{\sinh(t) \over 4} \sqrt{-1+2 t \coth(t)-{t^2 \over \sinh^2(t)}} - \xi^2 \right]^{-1}\ . \end{equation}
This is again a total derivative
\begin{equation}
{\dd \over \dd t}\left(-{1 \over 3}\,\xi^3 + {1 \over 4}\,\xi \sinh(t) \sqrt{-1+2 t \coth(t)-{t^2 \over \sinh^2(t)}}\right) =0\ ,
\end{equation}
with three solutions (for zero integration constant), $\xi=0$ and
\begin{equation}
\xi =  \pm{\sqrt{3} \over 2}\,\sqrt{\sinh(t)}\, \left(-1+2 t \coth(t)-{t^2 \over \sinh^2(t)}\right)^{1/ 4} = \pm{\sqrt{3} \over 2}\,\sqrt{\sinh(t)}\,e^{g/2}\ ,
\end{equation}
where the functional form of $e^g$ for the CVMN background can be read off from the last equality. 
Evaluating the Lagrangian (\ref{DBILag}) one finds for $\xi=0$
\begin{equation}
e^{-\phi}\sqrt{\mathrm{det}(G+\cF)} = {1 \over 8}\,\sinh(t)\,e^{g}\ ,
\end{equation}
and for $\xi = \pm\sqrt{3}\,e^x$
\begin{equation}
e^{-\phi}\sqrt{\mathrm{det}(G+\cF)} = {1 \over 4} \sinh(t) e^g \left( 1 + 3t^2 e^{-2g} \right)\ .
\end{equation}
For all three solutions the action clearly diverges exponentially in $t$ as $t \rightarrow \infty$
(this corresponds to a power divergence in $r$).
Therefore, the Euclidean D5-brane cannot be interpreted in terms of baryonic condensates.
This is in agreement with the fact that the CVMN solution does not belong to the
baryonic branch of the cascading gauge theory: its UV asymptotics are completely different from those
that define the cascading theories.

\section{Appendix: D7-Brane on the Baryonic Branch}
In this section we will briefly discuss the case of the D7-brane.
The first order equation (\ref{xiprime2}) with $\mathfrak{g}$ given by (\ref{g7}) can be rewritten in a form similar to
(\ref{FirstOrderEq})
\bea
\label{FirstOrderEqD7}
\xi'={e^g\over v}\nonumber &&\left[{\over}2h_2 a\sinh^2(t)(\xi+\chi)-(1+a\cosh(t))\left((\xi+\chi)^2-e^{2x}+h_2^2\sinh^2(t)\right)\right] \times\\
&&\Big[{\over}e^{2g}h_2\sinh(t)[1-e^{-2x}((\xi+\chi)^2-h_2^2\sinh(t)^2)] \nonumber
\\ &&-2 \sinh(t)(1+a\cosh(t))[a(\xi+\chi)
-h_2(1+a\cosh(t))]\,\Big]^{-1}\ .
\eea
Similarly to (\ref{D5_firstorder}), the $\xi$-dependent part of this equation can be represented as a total derivative
\bea \label{FirstOrderEqD7Int}
{d\over dt}&\bigg[&{1\over 3}(\xi+\chi)^3+{e^{2x} a\, (1+a\cosh(t))\over h_2 e^{2g}}(\xi+\chi)^2  \\ \nonumber
&-& [e^{2x} + h_2^2 \sinh^2(t) + 2e^{2x}(1+a\cosh(t))^2e^{-2g}](\xi+\chi)\bigg]
\\
= \nonumber
&-& {e^{2x} (1+a\cosh(t))\over v h_2 \sinh(t)
e^{g}}[e^{2x} - h_2^2 \sinh^2(t)] \\ \nonumber &-&[e^{2x} + h_2^2 \sinh^2(t) + 2e^{2x}(1+a\cosh(t))^2e^{-2g}]\chi'\ .
\eea
In analogy to (\ref{SInt}) the DBI action for D7 can be represented as the sum of a polynomial in $\xi$
and a $\xi$-independent integral
\bea \label{SD7}
S_{D7} = U\bigg|&-&{1\over 3}\tilde{\xi}^3+{a h_2 \sinh^2(t)\over 1+a\cosh(t)}\tilde{\xi}^2+
\left(e^{2x}-h_2^2\sinh^2(t)\right)\tilde{\xi}  \\ \nonumber
&+&\int_0^t\bigg[{h_2 \sinh(t) e^g \over v (1+a\cosh(t))}[e^{2x} + h_2^2 \sinh^2(t)] \\ \nonumber &+& {2 e^{2x} h_2 \sinh(t) (1+a\cosh(t)) \over v e^g} - [e^{2x} -
h_2^2 \sinh^2(t)]\chi'\bigg] dt\bigg|\ .
\eea
Interestingly, the coefficients of the characteristic cubic
polynomials in (\ref{FirstOrderEqD7Int}) and (\ref{SD7}) are the same ones
we encountered for the D5-brane, except that their roles are switched: $\mathfrak{C}, \mathfrak{D}$ and $\sigma$
appear in the differential equation for the gauge field while $\mathfrak{A}, \mathfrak{B}$ and $\rho$ appear in the action.

In the KS case (\ref{FirstOrderEqD7Int}) simplifies drastically and reduces to (compare with (\ref{KSSol}))
\bea
\xi^3=3 \left({(\sinh(t)\cosh(t)-t)^{2/3}h\over 16}+{(t\coth(t)-1)^2\over 4}\right)\xi\ ,
\eea
which has the trivial solution $\xi=0$ and a pair of non-zero solutions related to each other by the symmetry $\mathcal{I}$. From the asymptotic expansions (\ref{asymptotic}) of $\mathfrak{A}$ and $\rho$ it is then evident that the action (\ref{SD7}) will be exponentially divergent $\sim \mathcal{O}(e^{2t/3})$ for all three solutions.

\end{document}